\documentclass[referee]{aa}  
\bibpunct{(}{)}{;}{a}{}{,} 
\usepackage{graphicx}
\usepackage{subfigure} 
\usepackage{txfonts}
\usepackage{multirow}
\usepackage{enumerate}
\begin{document}

   \title{Galactic perturbations on the population of wide binary stars with exoplanets.}

   \author{J. A. Correa-Otto
          \and
          R. A. Gil-Hutton 
          }
   \institute
             {Grupo de Ciencias Planetarias, Dpto. de Geof\'isica y Astronom\'ia, Facultad de Ciencias Exactas, F\'isicas y Naturales, Universidad Nacional de San Juan - CONICET, Av. J. I. de la Roza 590 oeste, J5402DCS, Rivadavia, San Juan, Argentina\\
             \email{jorgecorreaotto@conicet.gov.ar} \\}

   \date{Received ; accepted }

 
  \abstract
   {}
   {The aim of this work is to study the dynamical effects of the Galaxy on binary star systems with physical and orbital characteristics similar to those of the population of known wide binary stars with exoplanets. As secondary goal we analyse the possible consequences on the stability of a hypothetical planetary system orbiting one of the stellar components. }
   {We numerically reproduced the temporal evolution of a sample of $3 \times 10^5$ binary star systems disturbed by the Galactic potential and passing stars in an environment similar to the solar neighbourhood. } 
   {Our results show that the dynamical evolution of the population of wide binary stars with exoplanets in the solar neighbourhood is modelled by the process of disruption of binary star systems induced by the Galaxy. We found that this process depends mainly on the separation between both stars, whereas it is almost independent of the initial orbital configuration. Moreover, our calculations are in agreement with the results of previous works regarding the indirect influence of the Galaxy on the stability of planetary systems in wide binary stars. However, the effects on the planetary region show a dependence on the initial configuration of binary stars.  Finally, we obtain an indirect test of the impulse approximation model for dynamical studies of binary star systems.}
   {}

   \titlerunning{Galactic perturbations on wide binary stars with exoplanets}
   \maketitle

%


\section{Introduction}\label{intro}

Binary star systems are generally classified as tight binary stars, for which the distance between the components is less than 1000 au, and wide binary stars (WBS), with mutual distance greater than 1000 au. Currently, there are 18 known WBS harbouring 28 exoplanets (Roell et al. 2012), which comprises $\sim 35 \%$ of the planets  detected in binary stars.

Exoplanets in WBS have been ignored in dynamical studies, probably because a distant stellar companion produces only very weak effects on the planetary evolution (Rabl \& Dvorak 1988; Holman \& Wiegert 1999; Andrade-Ines \& Michtchenko 2014). However, WBS are not isolated systems; they evolve in a Galactic environment under the influence of external perturbations, where passing stars and the tidal field of the Milky Way are the most important disturbers. Their effects on the dynamical evolution of WBS were studied in several works (Heggie 1975;
Bahcall et al. 1985; Jiang \& Tremaine 2010). Moreover, similar studies have also been applied in  external solar system dynamics research, in particular the Kuiper belt and Oort cloud (Duncan et al. 1987; Brunini 1995; Eggers \& Woolfson 1996; Levison \& Dones 2001; Fouchard et al. 2006; Kaib et al. 2011).

The discovery of structures similar to the Kuiper belt in other planetary systems (Moro-Martin et al. 2015; Kennedy et al. 2015) and the existence of exoplanets in WBS have stimulated investigations of the influence of the Galactic environment on extrasolar planetary systems. Recently, Kaib et al. (2013) have proposed that the apparent more eccentric orbits of exoplanets in WBS can be explained by such external perturbations.

Although the pioneering work of  Kaib et al. (2013) has shown interesting results, a complete portrait of the problem remains unknown. The dynamics of planets orbiting one of the WBS components and the effects of  the galactic environment define a complex problem with a large number of possible initial configurations and physical parameters. In fact,  Kaib et al. (2013) only studied 2600 possible binary star configurations, in a phase space of six dimension and two parameters (i.e. stellar masses), and considering only a uniform distribution on  $\log{a}$  and $e^2$.

Thus, even though we are mainly interested in studying the
evolution of planets
orbiting one of the stars of a WBS, since\ the Galaxy affects the planetary system indirectly through perturbations on the stellar companion and the masses of the planets are much lower than those of the stars, we can start studying the dynamical behaviour of the binary star system and then get information on the stability of the planetary system using the critical periastron criterion defined by  Kaib et al. (2013).

Additionally, both  Kaib et al. (2013) and other authors have considered a model of impulse approximation to simulate the stellar passage (Levison \& Dones 2001; Zakamska \& Tremaine
2004; Rickman et al. 2008; Jiang \& Tremaine 2010; Kaib et al. 2011; Kaib \& Raymond 2014). Although Yabushita et al. (1982), Scholl et al. (1982), Dybczynski (1994), and Eggers \& Woolfson (1996) have demonstrated a good agreement of that model with the numerical simulations for the case of a restricted three-body problem (e.g. Sun, comet, and star),  it is unclear if it will work in the general three-body problem. This is an important topic because in a restricted three-body problem the two massive bodies describe a parabolic orbit and there is no chance for a mutual capture in an elliptical orbit. Instead, in a general three-body problem we initially have a binary system whose centre of mass moves in a hyperbolic orbit with the third body, but in the instance of a close approach the dynamical evolution becomes chaotic and could occur a stellar reconfiguration with the formation of a new binary star system or even a triple star system. Despite the increase of the computing capacity in recent years, we did not find a test for the impulse approximation model in systems with more than two massive bodies in the literature.

The aim of this paper is to study the dynamical evolution of a synthetic sample of  WBS under the effects of the gravitational potential of the Galaxy and passing stars to improve our understanding of the dynamic effects of the Galaxy on wide binary stars and to study the indirect effects on planetary systems. In Section \ref{model} we describe the numerical methods employed and the choice of the initial configurations. In section \ref{result} we present our results. Finally, conclusions close the paper in Section \ref{conclu}.

\section{Numerical methods}\label{model}  

\subsection{Wide binary star systems}\label{system}

Table \ref{table0} shows the list of binary star systems harbouring  exoplanets with projected separations ($\rho$) greater than 800 au (Roell et al. 2012; Kaib et al. 2013). We used these data to make a synthetic population of $10^5$ binary star systems and follow their temporal evolution during a period of 10 Gyr, i.e. the estimated age of the thin disk of the Milky Way. The masses of the primary and secondary stars of each WBS, $m_1$ and $m_2$, are taken at random from Table \ref{table0} and we assume that the planets are always orbiting the primary star.

\begin{table}
\caption{Physical parameters of the known WBS with exoplanets. The stellar masses and their projected separation ($\rho$) are in the third, fourth, and fifth columns, respectively. The index $1$ in the masses indicates the member of the binary system harbouring the planetary system.  The mass of the secondary stars indicated with `RD' (red dwarf) corresponds to the range 0.08 $-$ 0.5 M$_\odot$. The second and last columns indicate the number of exoplanets  and semi-major axis ($a_p$) of the most external planet orbiting the main star ($m_1$).}
\label{table0}
\centering
\begin{tabular}{c c c c c c  }
\hline\hline
  WBS     & $N_P$ & $m_1$     &  $m_2$     & $\rho$ & $a_p$  \\
  Name    &       & m$_\odot$ &  m$_\odot$ & au     & au  \\
\hline
GJ 676 A  &     1     & 0.71      & $-$        &        800 &    1.82    \\
HD 142022 &     1     & 0.9       &     $-$        & 820    &    2.93      \\
11 Comae  &     1     & 2.04      &     $-$            & 995    &        1.29       \\
HD 11964  &     2     & 1.08      &             0.67   & 1044   &       3.34           \\
55 Cancri &     5     & 0.905     &     0.13       & 1065   &   5.74              \\
HD 80606  &     1     & 0.958     & $-$        & 1197   &       0.453      \\
HD 204941 &     1     & 0.74      & $-$        &   1447 &       2.56    \\
HAT-P-1   &   1   &  1.15     & 1.16       &   1557 &   0.06              \\
HD 101930 &     1     & 0.74      &     $-$            &   2227 &       0.3    \\
HD 7449   &   1   &1.05       & $-$            &   2348 &       4.96   \\
HD 190360 &     2     & 0.983     & 0.2        &   3293 &       3.92   \\
HD 213240 &     1     & 1.14      & 0.14       &   3898 &       1.92    \\
HD 147513 &     1     & 1.072     & $-$        &   4460 &       1.32        \\
HD 222582 &     1     & 0.965     & RD         &   4595 &       1.35      \\
XO-2      &   1   & 0.97      & $-$            &   4619 &       0.04      \\
HD 125612 &     2     & 1.1           & RD             & 5990   & 4.2     \\
HD 20781  &     2     & 0.84      & 0.969      &   9133 &       1.38      \\
HD 20782  &     1     & 0.969     & 0.84       &   9133 &       1.38      \\
HD 38529  &     2     & 1.34      & RD         &  11951 &       3.695     \\
\hline
\end{tabular}
\end{table}

We considered a Cartesian astrocentric coordinate system $(x, y, z)$ with origin in the main star ($m_1$). This system is at a distance $R_g$ from the Galactic centre and corotates with the Galaxy. The $z-axis$ is perpendicular to the Galactic plane and points towards the south galactic pole, the $y-axis$ points in the direction of the Galactic rotation, and the $x-axis$  points radially outwards from the Galactic centre. In such a system, the secondary star ($m_2$) evolves around the primary ($m_1$) in an orbit of size and shape defined by a semi-major axis $a$ and eccentricity $e$. For the angular orbital elements we assume an isotropic distribution and the inclination of the orbit is defined with respect to the Galactic plane.

 \begin{figure}
\centering
\subfigure{\includegraphics[width=0.7\textwidth]{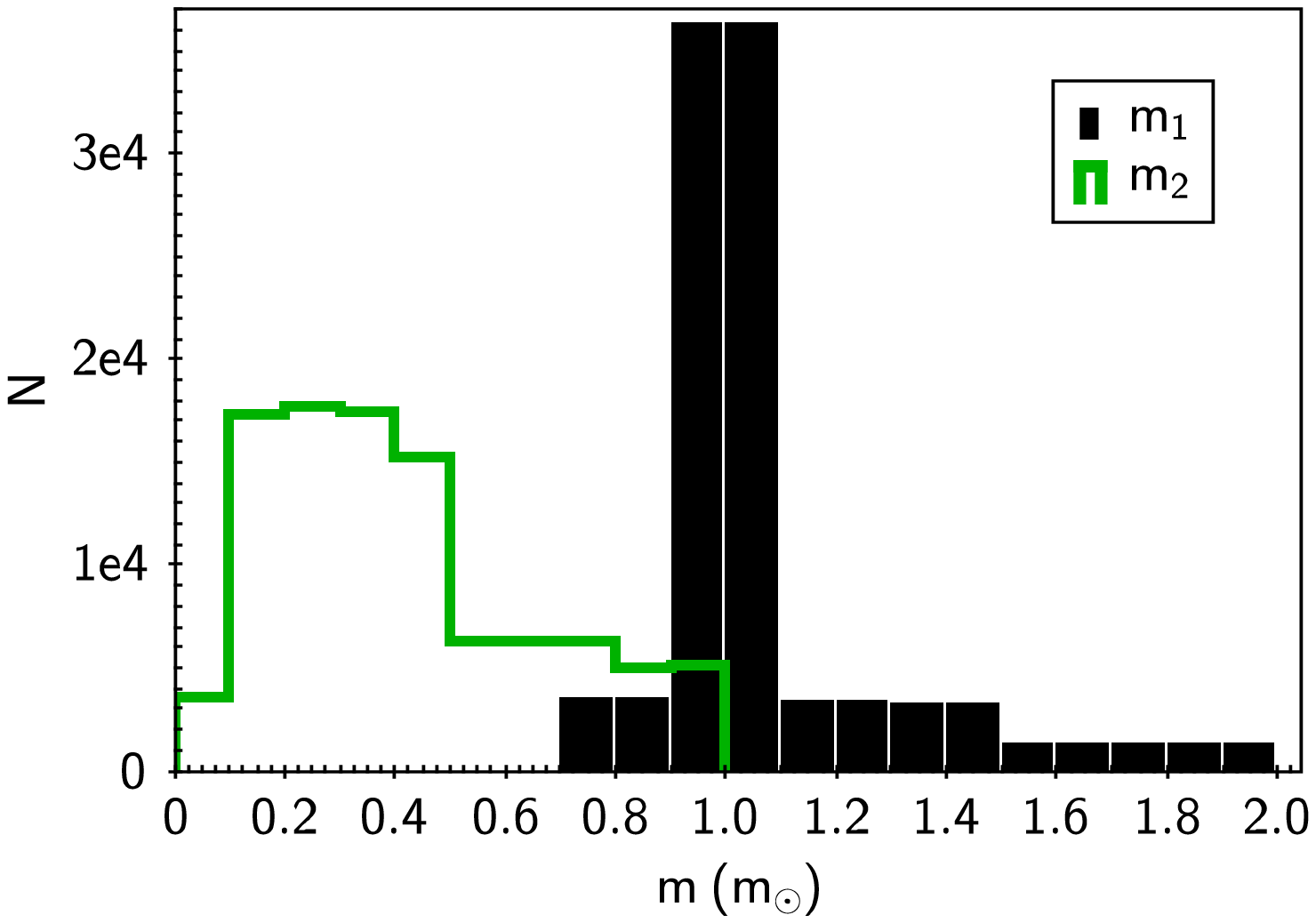}}
\subfigure{\includegraphics[width=0.7\textwidth]{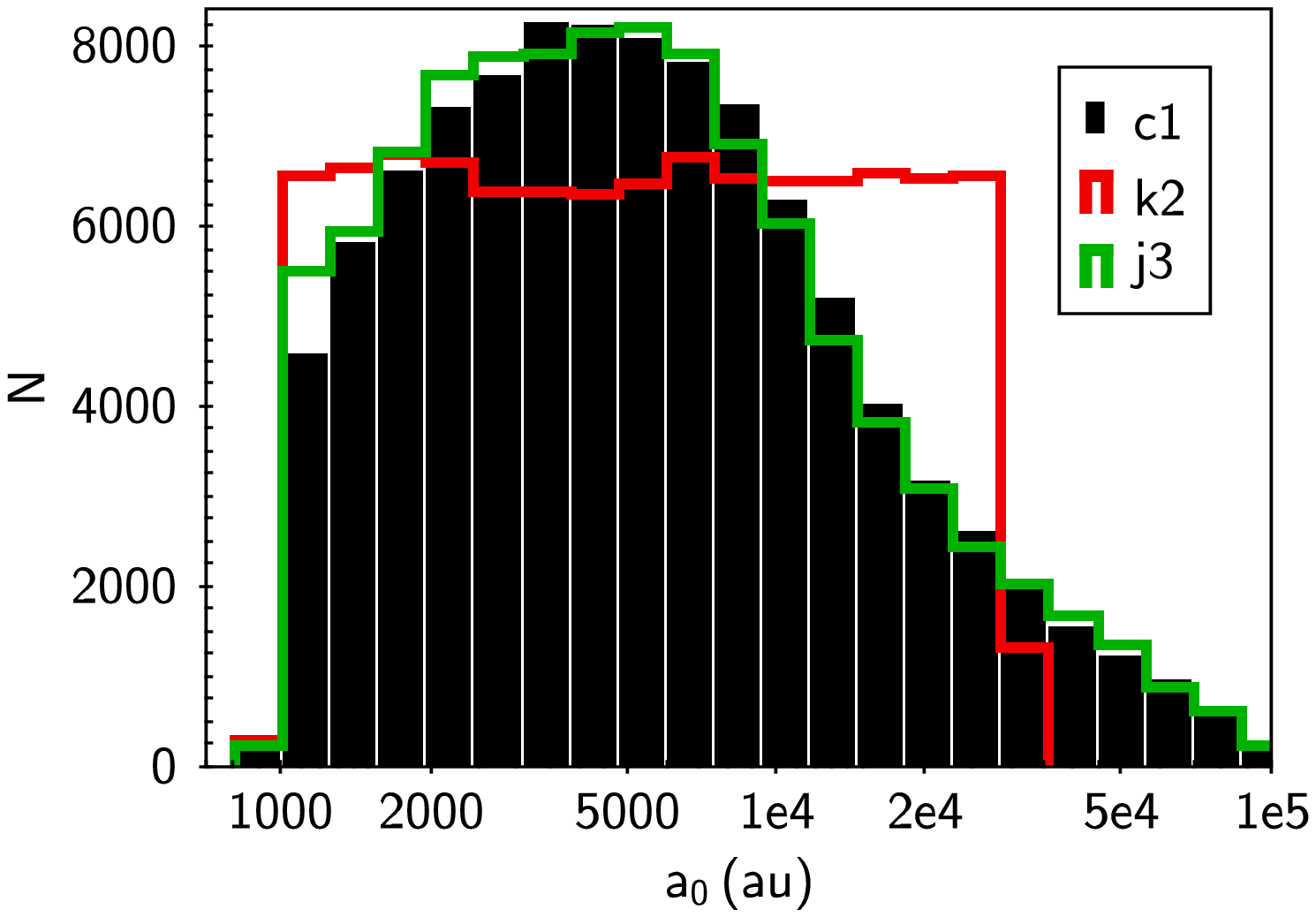}}
\subfigure{\includegraphics[width=0.7\textwidth]{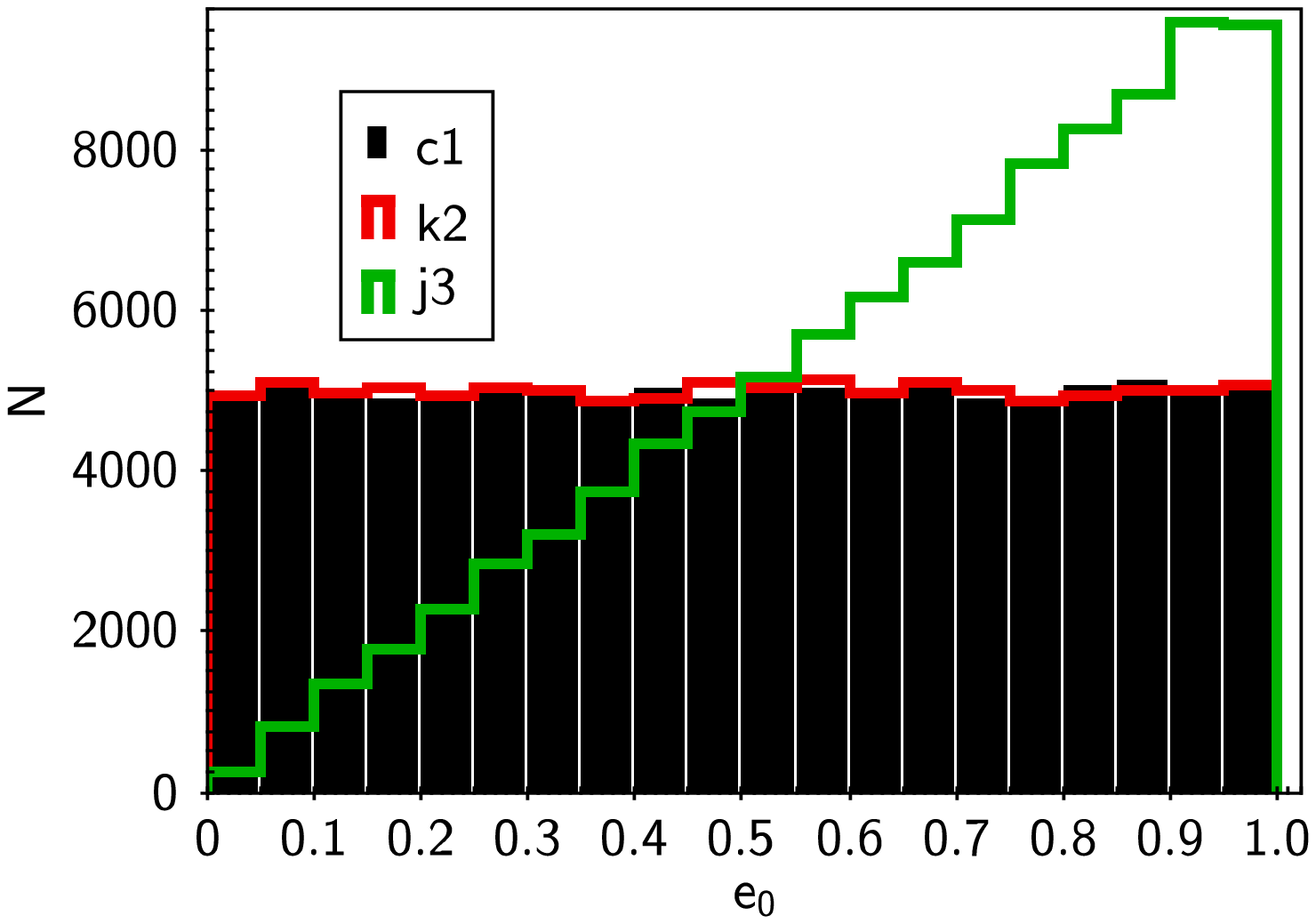}}
\caption{{\small Parameters and initial configuration of the three sets. Top panel: Initial distribution of masses for the three sets of WBS is shown. The black and open green histograms correspond to the main star ($m_1$) and the secondary star ($m_2$), respectively. The middle and bottom panels show the initial distribution of the semi-major axis and eccentricity for the three sets of WBS: \textbf{c1}, \textbf{k2,} and \textbf{j3} in black, open red, and open green histograms, respectively.}}
\label{fig0}
\end{figure}

The initial distributions of the semi-major axis and eccentricity are difficult to determine because we only know the projected distance $\rho$ (fifth column, Table \ref{table0}) between the members of the WBS. So, we define three different combinations of both elements to be assigned to three synthetic population sets. We define a first set of  $10^5$ synthetic WBS (called \textbf{c1}) with an isotropic distribution of $e$. Then, assuming that the line between the two stars has a random angle to the line of sight, we can estimate the distribution of the real separation between the binary components; this distribution was used together with the eccentricity and mean anomaly to obtain the distribution of semi-major axis, taking as lower and upper limits semi-major axes of 1000 au and 100000 au, respectively.

We obtained the second set of $10^5$ synthetic WBS (called  \textbf{k2}) by keeping the distribution in $e$ of the previous set, but considering a uniform distribution in $log \, a$, with $a \in (1000,30000)$ au, which is similar to that used by Kaib et al. (2013). Finally, for the third set (\textbf{j3}), we considered an initial random distribution in $e^2$, following Jiang \& Tremaine (2010) and Kaib et al. (2013), and found the semi-major axis from $e$ and the projected distance as in the \textbf{c1} set.

The initial configuration of the three sets is shown in Fig. \ref{fig0}, where the mass distribution used  is in the top panel, while in the middle and bottom panels are shown the distribution of $a$ and $e$ in black, open red, and open green histograms for the \textbf{c1}, \textbf{k2}, and \textbf{j3} sets, respectively. The purpose of using these three sets is to take into account the different initial configurations analysed in the literature to improve our understanding of the Galactic effects on the population of WBS with exoplanets. The total sample of $3 \times 10^5$ WBS is two orders of magnitude larger than that of Kaib et al. (2013).

The temporal evolution of the  $3 \times 10^5$ WBS was solved numerically by integrating the exact equations of motion using a Bulirsch-Stoer code with adopted accuracy of $10^{-13}$. During the simulation the WBS are affected by the external influence of the tidal field of the Milky Way and passing stars. We considered the solar neighbourhood as the Galactic environment for our simulations because $\sim$ 80 \% of the WBS with exoplanets are at distances less than 50 pc from the Sun, but it is important to mention that these objects possibly spend much time in denser regions closer to the Galactic centre (Sellwood \& Binney 2002; Roskar et al. 2008; Kaib et al. 2011) and the typical perturbations would be more powerful than we considered in our scenario.

Finally, we repeated
the simulation of the  \textbf{c1} set considering each perturbation separately to improve our comprehension about the external perturbations acting, their interactions, and the effects produced on the WBS. Then, we define two subsets that are identical to the \textbf{c1} set. But for the first subset, called \textbf{c1a}, we only considered the effect of the potential of the Galaxy, while for the second subset, \textbf{c1b}, we only took the influence of stellar passages into account.

\subsection{Galactic tidal field}

We use the Hill approximation (Heggie 2001; Binney \& Tremaine 2008) to describe the motion of a WBS in the Galaxy and assume that it is at $R_g=$ 8 kpc from the Galactic centre. Assuming the non-inertial Cartesian astrocentric coordinate system described in the previous Section and a symmetric potential on the plane $z = 0$, the corresponding equations of motion are

\begin{equation}
\begin{array}{cclccl}
\ddot{x} &=& - \dfrac{\mu x}{r^{3}} + 2 \Omega_G  \dot{y} + 4 \Omega_G A_G x \,  \rm{,}\\ 
\\
\dot{y} &=& - \dfrac{\mu y}{r^{3}} - 2 \Omega_G  \dot{x}  \,  \rm{,}\\ 
\\
\ddot{z} &=& - \dfrac{\mu z}{r^{3}} - \nu_G^2 z  \,  \rm{,}\\ 
\label{eq1}
\end{array}
\end{equation}

\noindent where, $r=\sqrt{x^2+y^2+z^2}$, $\mu= \mathcal{G}  (m_2 + m_1)$, $\mathcal{G}$ is the Gravitational constant, and $x$, $y$, $z$, $\dot{x}$, $\dot{y}$, $\dot{z}$ are the components of the astrocentric position and velocity of the secondary body $m_2$ around $m_1$. Moreover, $\Omega_G$, $A_G$, and $\nu_G$ are the angular speed of the Galaxy, the Oort constant, and  the frequency for small oscillations in $z$, respectively. At the approximated distance of the Sun from the Galaxy centre (i.e. $R_g = 8 $ kpc ) their values are

\begin{equation}
\begin{array}{ccl}
\Omega_G &=& 3.017 \times 10^{-8} \,  yr^{-1}\,  \rm{,} \\ 
A_G &=& 1.513 \times 10^{-8} \,  yr^{-1}\,  \rm{,} \\ 
\nu_G &=& 7.258 \times 10^{-8} \,  yr^{-1} \,  \rm{.} \label{eq2}
\end{array}
\end{equation} 

\noindent For more details about the deduction of Eq. (\ref{eq1}) see Jiang \& Tremaine (2010).

\subsection{Stellar perturbations}

The effect of a passing star is included as the encounter of a WBS with a third star, which we identified as $m_3$. Usually, this three-body interaction is computed with the model of impulse approximation (Rickman 1976). However, in spite of the high relative velocities between the stars on the solar neighbourhood, we solve numerically by a direct integration of a three-body problem with the additional perturbation of the Galactic potential. 

For a relative velocity of 40 km s$^{-1}$ and a number density of stars of 0.05 pc$^{-3}$ in the solar neighbourhood, the total number of stellar encounters with impact parameter less than $q$ during the total time of integration ($T=$ 10 Gyr) is defined as (Jiang \& Tremaine 2010; Brunini \& Fernández 1996)

\begin{equation}
N = 800 \, \left( \dfrac{q}{0.1 \rm{pc}} \right) ^2   \,  \rm{.} \label{eq3}
\end{equation}

\noindent We consider for all the sets a maximum impact parameter of $q_M \sim$ 1 pc.  This value corresponds to a total of $N = 8 \times 10^4$  stellar passages of a background star in the vicinity of a WBS, occurring at random times during the total time of integration. Each stellar passage was generated following the method of Rickman et al. (2008) and Kaib \& Raymond (2014). The mass of the third star ($m_3$) is selected from the mass-luminosity function in the solar neighbourhood (Reid et al. 2002; Ninkovic \& Trajkovska 2006) and the initial relative velocity for the encounter is taken from the velocity dispersion of nearby stars available in the Hipparcos data (Garcia-Sanchez et al. 2001), which is a function of the stellar masses. 

For each one of the WBS on the three sets, the $8 \times 10^4$ stellar passages are randomly distributed along the total time of integration. Then, in our simulations each one of the $3 \times 10^5$ WBS has a different set of stellar encounters.

\section{Results}\label{result}

The main influence of the Galaxy on the WBS are twofold: first, the disruption of the binary system, which produces gaps in the orbital elements distributions of the three sets; and second, the changes in the orbital configuration of the surviving WBS, which modify the orbital element distributions of the population in each set.


 \begin{figure}
\centering
\subfigure{\includegraphics[width=0.7\textwidth]{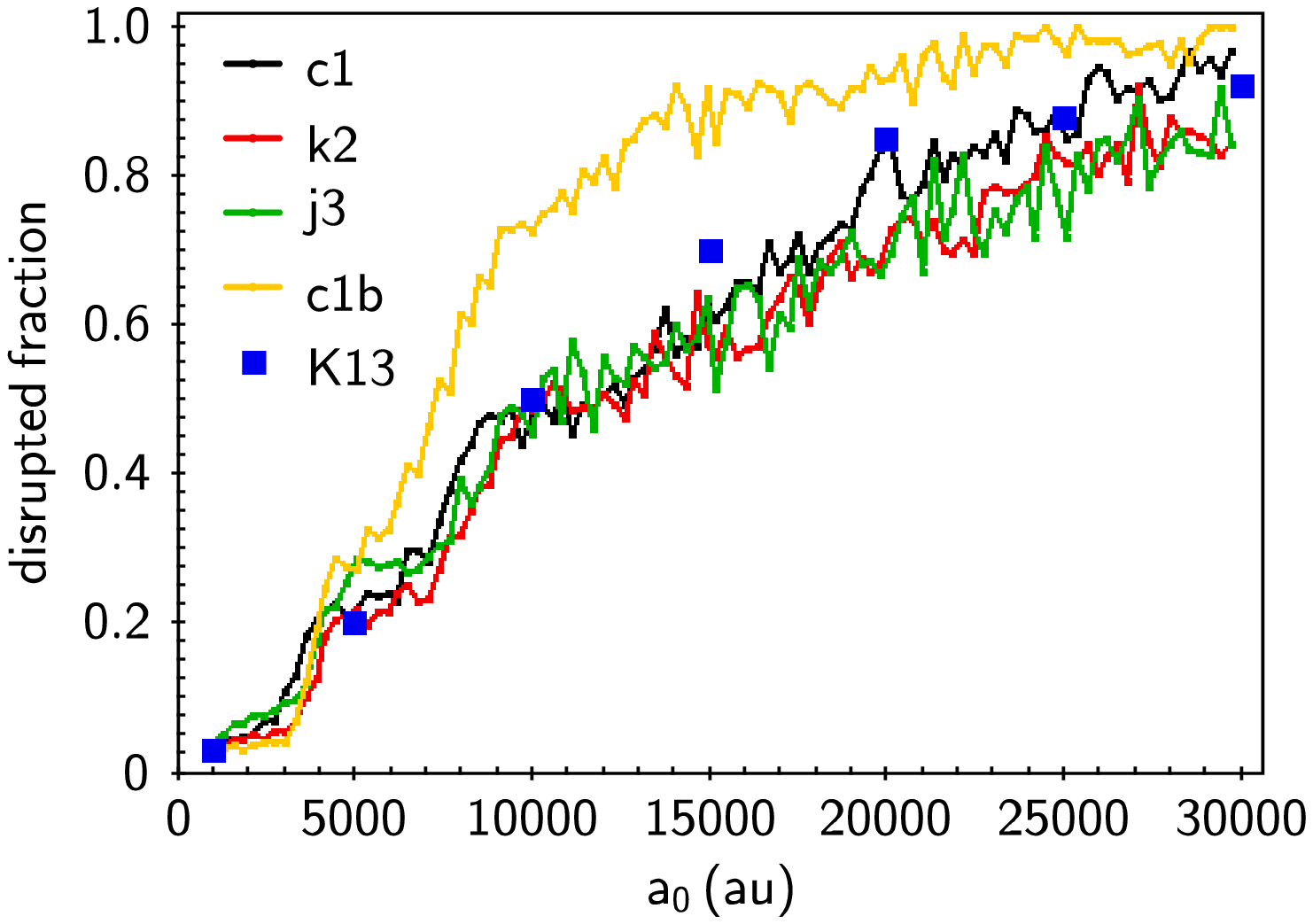}}
\subfigure{\includegraphics[width=0.7\textwidth]{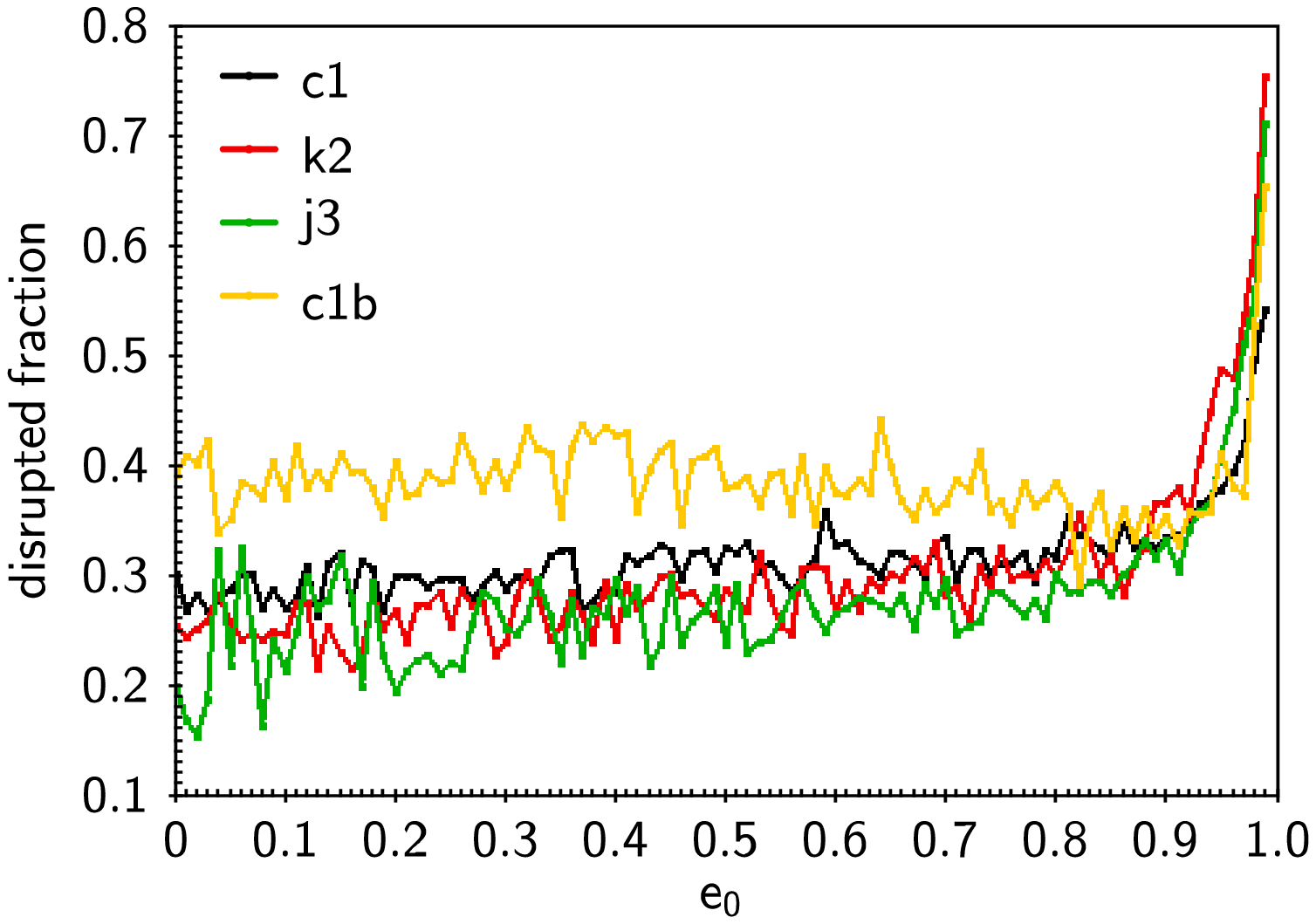}}
\subfigure{\includegraphics[width=0.7\textwidth]{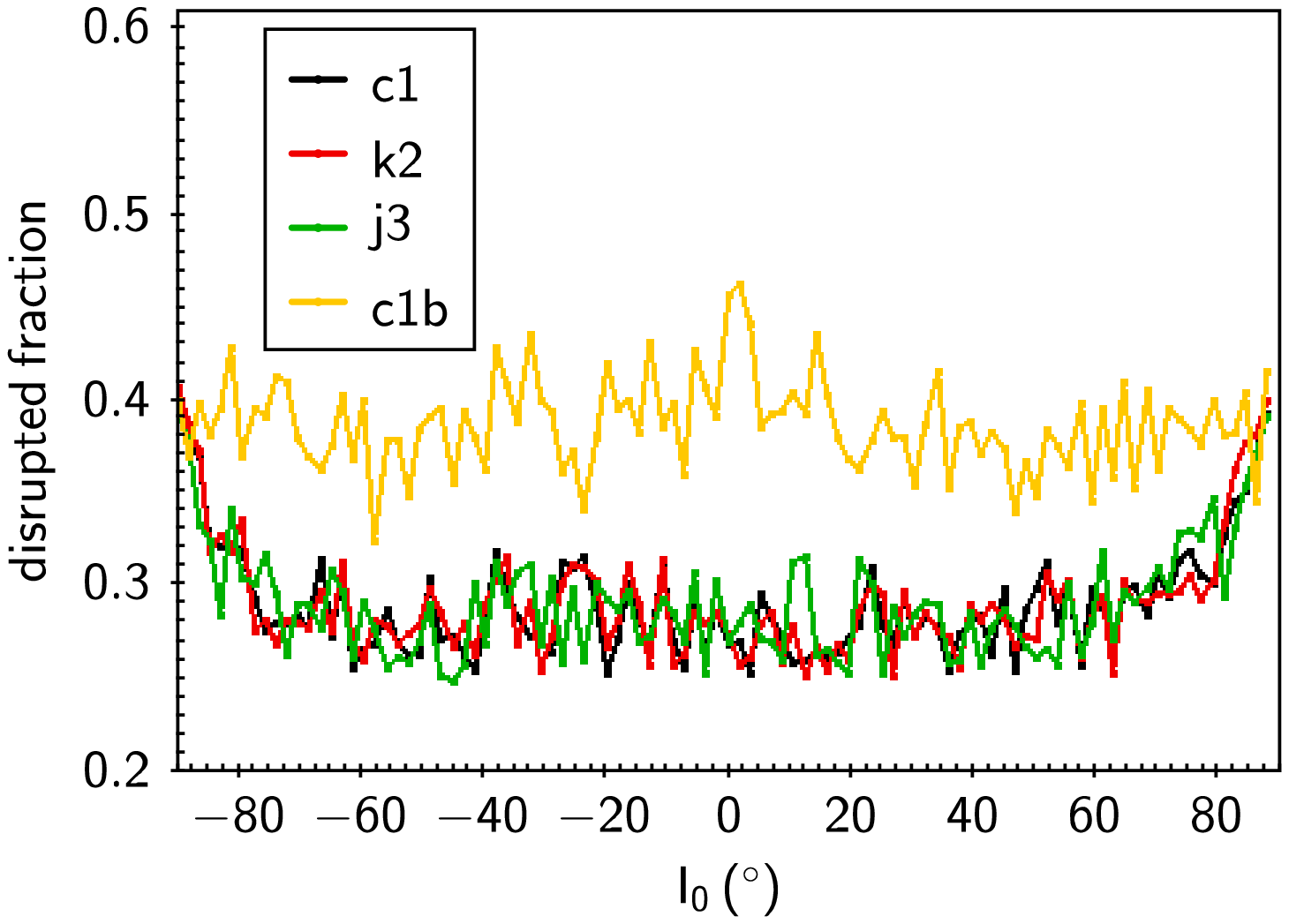}}
\caption{{\small Top panel: Fraction of disrupted WBS in the initial range $a_0 \in$ (1000, 30000) au.  The blue squares are results taken from the Figure S8 of Kaib et al. (2013) included for comparison. Middle and bottom panel: Fraction of disrupted WBS as a function of the initial eccentricity and inclination, respectively. In the three graphs the sets \textbf{c1}, \textbf{c1b}, \textbf{k2,} and \textbf{j3} are shown in black, yellow, red, and green, respectively.}}
\label{fig1}
\end{figure}

We considered that a WBS is disrupted when the main star loses its companion or the distance between their components is less than 2 stellar radii (Kaib \& Raymond 2014; Correa-Otto et al. 2017). Then, the disruption conditions are $e \geq$ 1 and $a (1-e) \leq 2 R_*$ for each case, respectively. We obtain that $\sim$ 28 \% of the WBS in each sample (i.e. \textbf{c1}, \textbf{k2,} and \textbf{j3}) are disrupted by the external perturbations. This is an interesting result because it seems to indicate that the rate of disruption of WBS in the solar neighbourhood is independent of the initial distribution of the orbital configuration. On the other hand, for the set where we only take into account the influence of stellar passages (\textbf{c1b}) we obtain a 40 \% of disrupted WBS. This result indicates that the stellar passages would have a destabilizing effect on the WBS population, but its intensity would be reduced when it acts in combination with the Galactic potential. Moreover, in the sample \textbf{c1a} all the disrupted WBS correspond to stellar collisions indicating that the Galactic potential alone is not able to ionize the WBS. For all the sets the probability of a close encounter between $m_1$ and $m_2$ is statistically negligible (i.e. $\ll 1 $\%). Finally, we found other stellar configurations, such as a new binary system between $m_1$ and $m_3$ or a triple star system, but their percentage of occurrence is very small ($\ll$ 1 \%)  making it possible to ignore these configurations without modifying our statistical results.

Our results are in agreement with those of Kaib et al. (2013) who found a $\sim$ 25 \% of disrupted systems in a sample with a distribution in $a$ and $e$ similar to our sets \textbf{k2} and \textbf{j3}, respectively. This is an interesting result because these authors used the model of impulse approximation  for the stellar passages, while we solved the complete three-body problem. Then, the impulse approximation seems to be sufficiently good to study the population of WBS with exoplanets.

In order to determine the importance of the orbital configuration of WBS during their  dynamic evolution on the Galaxy, we analyse the disruption process as function of the initial distribution of the elements that define the extension, form, and orientation of the orbit (i.e. $a$, $e,$ and $I$). We develop this study following the technique of Kaib et al. (2013), which consists in calculating the proportion of disrupted systems along  an orbital element. Additionally, this method allow us to compare our results with that work. For the semi-major axis we limit our analyses to $a \in$ (1000, 30000) au in our sample \textbf{k2} because this is the range used by  Kaib et al. (2013) and also because the other two sets have a small percentage of WBS with $a>$ 30000 au. For each sample we divide the complete range in 100 intervals and count the quantity of WBS in each one, and the quantity of disrupted WBS. The percentages of disrupted WBS per bin are shown in Figure \ref{fig1}, top panel, where the three samples are shown in black (\textbf{c1}), red  (\textbf{k2}), and green (\textbf{j3}), and some results of \cite{kaib13} are indicated by blue squares. We can see an agreement between our results and that work, which indicates that the probability of the rupture of a WBS increases with the separation between the components of the binary star. On the other hand, the subset \textbf{c1b} (yellow line) confirms our previous results that  the dissociation process of binary star systems in the solar neighbourhood is dominated by stellar passages; however, the combination with the effect of the Galactic potential reduces the number of disrupted WBS.

The middle and bottom panels of Figure \ref{fig1} also show the fraction of disrupted WBS as functions of their initial eccentricity and inclination, where the sets are indicated with the same colours used in the top graph. The results for the three sets look similar and confirm the independence of the disruption process with the initial configuration of the WBS. The systems with $e>$ 0.9 have a high percentage of disruptions because the orbits of these objects are quasi-parabolic and the passage by the apocentre is very slow, facilitating its rupture. It is important to mention that the WBS with initially highly inclined orbits ($I > \pm$ 80$^\circ$) have a large percentage of disruptions in comparison with WBS in low and intermediate inclinations even when the effects of the Galactic tide and stellar passages are combined, reaching a percentage of disruptions similar to that of the \textbf{c1b} set.

Furthermore, our results show that the disruption process is independent of the WBS masses or, in other words, the disruption probability is approximately the same for any value of $m_1$ or $m_2$. Thus, our dynamical study is not affected by the poorly determined values of the secondary star masses. Additionally, an important consequence of this result is that the evolution of the WBS in the Galaxy does not change the distribution of their masses (Table \ref{table0}). Then, the initial and final mass distribution normalized with the total number of cases considered should be similar.

 \begin{figure}
\centering
\subfigure{\includegraphics[width=0.7\textwidth]{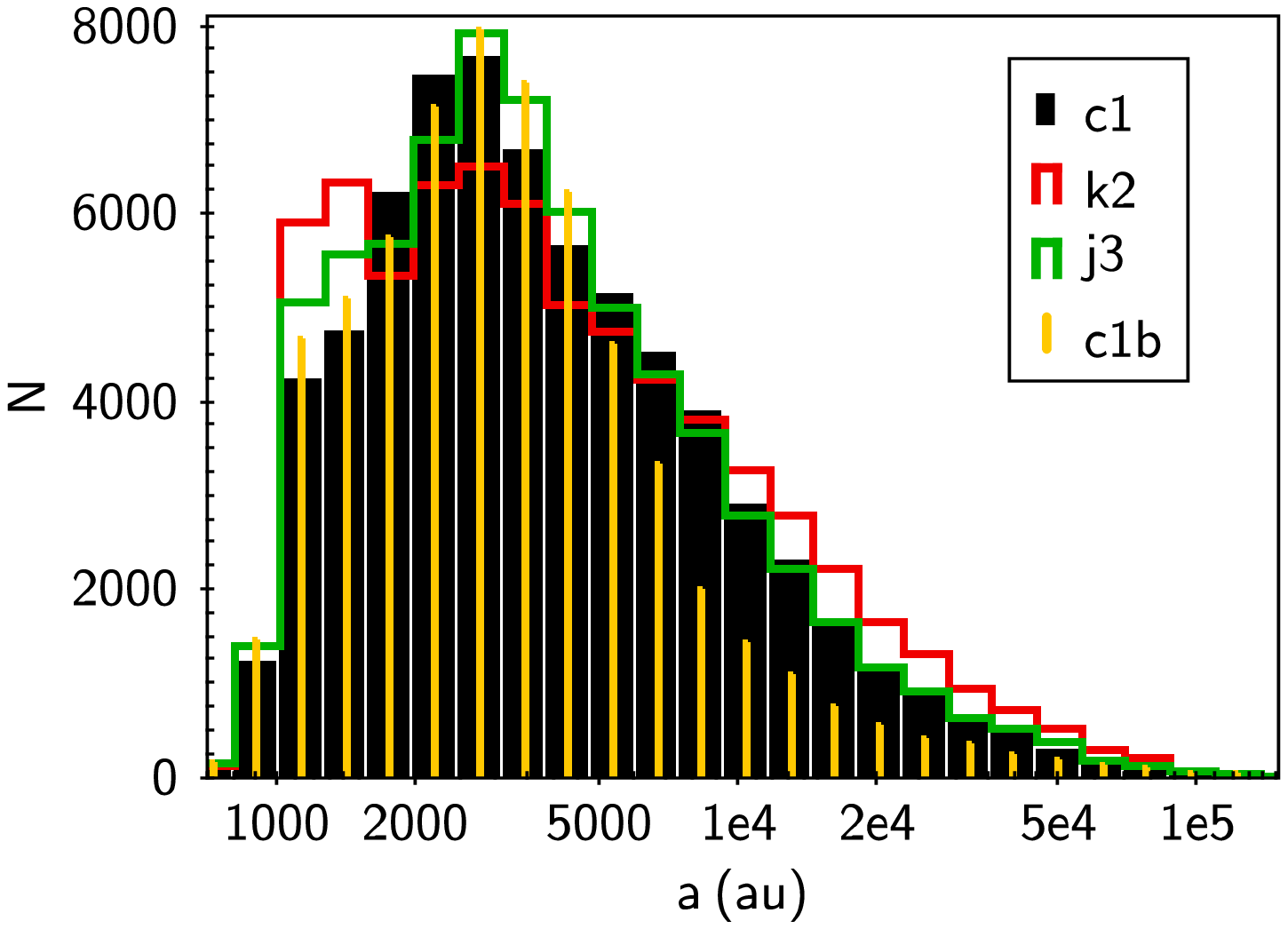}}
\subfigure{\includegraphics[width=0.7\textwidth]{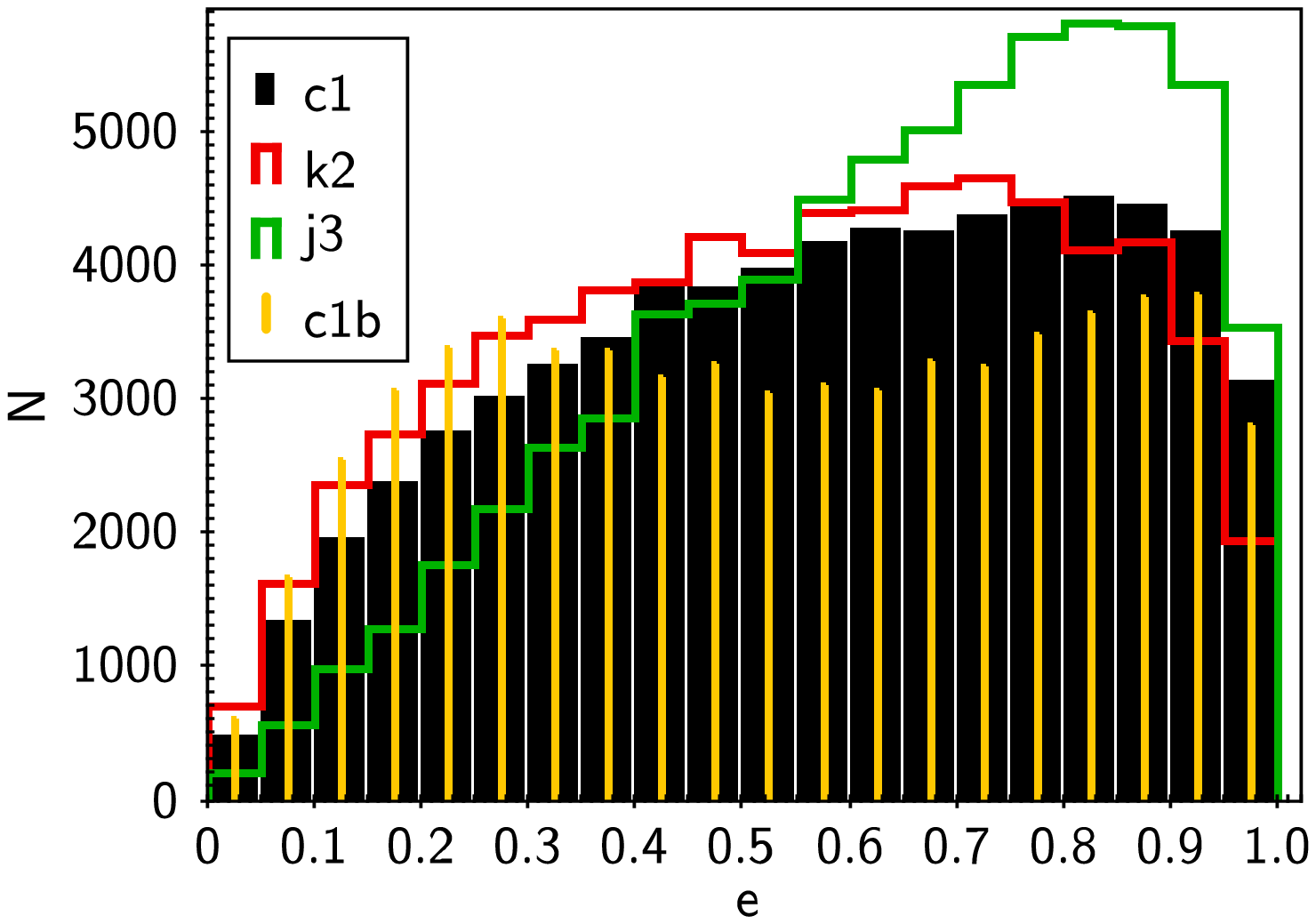}}
\subfigure{\includegraphics[width=0.7\textwidth]{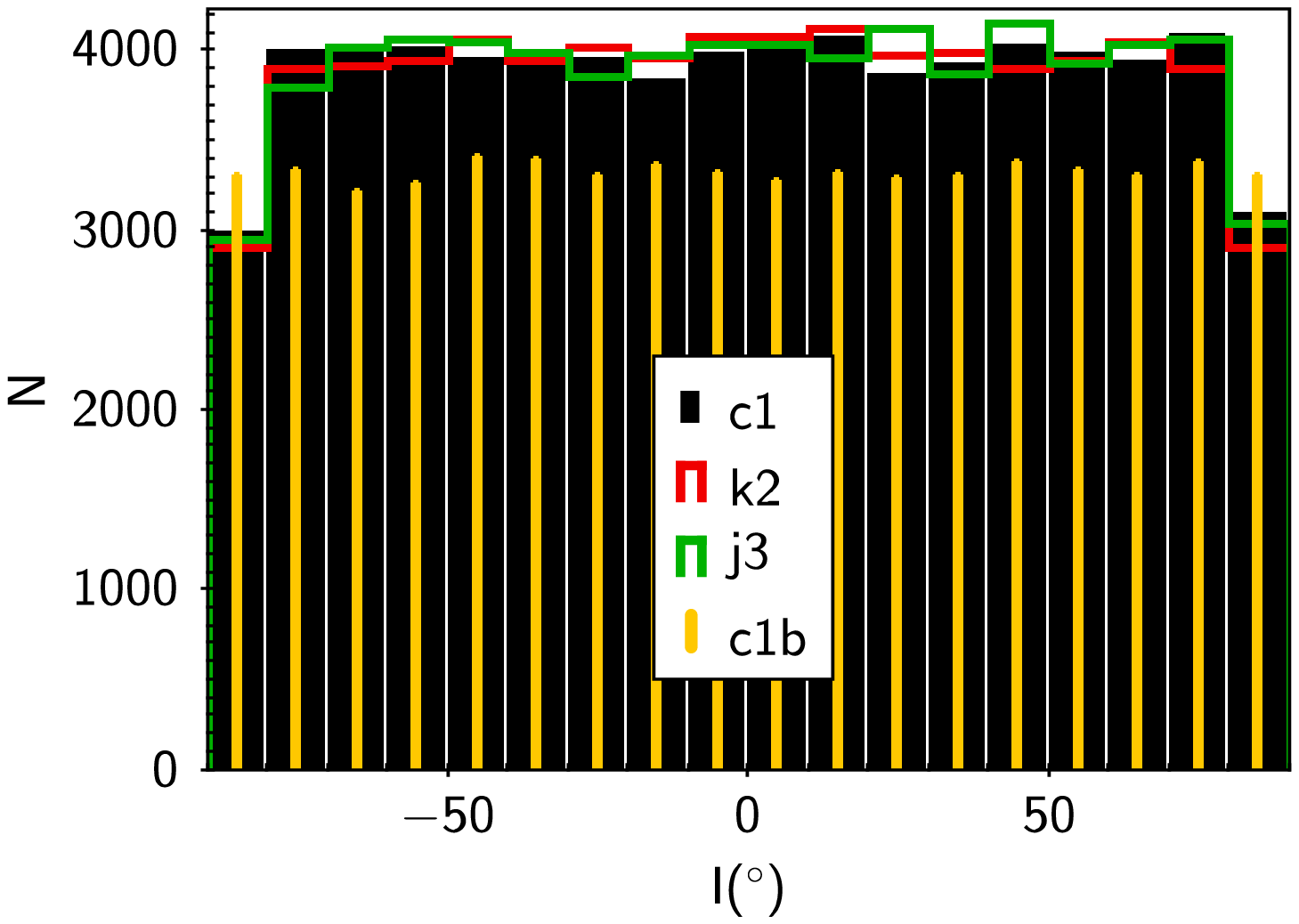}}
\caption{{\small Orbital distribution for the sets \textbf{c1}, \textbf{k2,} and \textbf{j3} after $T=$ 10 Gyr of evolution. The three samples are indicated with histograms in black, red, and green, respectively, while the subset \textbf{c1b} is indicated with a yellow spike histogram. The top, middle, and bottom panels correspond to the orbital elements $a$, $e,$ and $I$, respectively.  }}
\label{fig2}
\end{figure}

On the other hand, the dynamical influence of the Galactic environment in the surviving synthetic population of WBS is analysed considering the final distribution of the orbital elements $a$, $e,$ and $I$ of each set. Figure \ref{fig2} shows our results for the samples \textbf{c1}, \textbf{c1b}, \textbf{k2,} and  \textbf{j3}.  As in the disruption dynamical process, the three populations end with similar orbital characteristics, which confirms the independence with respect to the initial configuration of WBS. The final distribution of the semi-major axis is in the top panel, where we plot it in log-scale because of the high percentage of disrupted WBS with large separations between the stars. We observe for the three sets a maximum at $\sim$ 2500 au, which seems to be a result of a combination of the great slope for the initial distribution at small $a$ (see Fig. \ref{fig0}) and the disruptive process of the effects of the Galaxy for high values of $a$. This maximum at $\sim$ 2500 au also appears in the subset \textbf{c1b}, which seems to indicate a limit for the external effects on WBS and provides a dynamical explanation for the empirical limit of 1000 au between tight and wide binary star systems.

The middle panel of Fig. \ref{fig2}, shows the final distribution of the eccentricity for the three sets. We can see a maximum at high eccentricities, close to a value of 0.8. Since this maximum is not present in the \textbf{c1b} subset, it is possible that the evolution of the eccentricity could be a consequence of the interaction of the two phenomena, the stellar passages, and the Galactic potential, but in any case this is a marginal result.

In any case, the final distribution of $e$ represents an important dynamical result. For the samples \textbf{c1} and \textbf{k2} we assume that the binary stars do not have a preferred eccentricity when they are born, but the effects of the Galaxy change the uniform distribution and define an excited or hot population. For the \textbf{j3} set the initial distribution of $e$ corresponds to an initially hot population, but the dynamical evolution disrupts WBS with very high values of $e$, and the final distribution is similar to those of the other two samples. Then, our results show that the influence of the  Galaxy changes the form of the WBS orbits and reaches a final distribution with a high eccentricity maximum with independence of the initial distribution of $e$.


The final distribution of $I$ (bottom panel of  Fig. \ref{fig2}) is modelled by the disruption process. There is a gap  in the region of $\pm$ 90 degrees, which as explained  seems to be consequence of the dynamical effects of the combined phenomena of Galactic potential and stellar passages.  Hence, most of the population of WBS in the solar neighbourhood would have orbits with low and medium inclinations.

\subsection{Influence of the Galactic perturbations: Sets \textbf{c1} and \textbf{c1b}}

 In order to understand the effects of the Galactic potential and stellar passages it is necessary to perform a better analysis to explain the differences between the results obtained for sets \textbf{c1} and \textbf{c1b}. Such differences show an important dependence on the separation of the pair and the inclination of the orbit. Hence, we have to explain these results.

For a binary star system disturbed by the Galactic tidal field there is a Jacobi or tidal radius, which is a stationary solution of the Eq. (\ref{eq1}), and it is defined by
\begin{equation}
r_J = \left[ \dfrac{\mu }{4 \Omega_g A_g} \right]  ^{1/3}\,  \rm{.} \label{eq4}
\end{equation}
Jiang \& Tremaine (2010) found that $r_J$ defines a characteristic scale for the distribution of binary stars systems with large separations. The existence of a stability limit for the separation of a binary star disturbed by the Galaxy can give us the wrong impression that we have to expect a higher disruption rate for the sample \textbf{c1}, which is the opposite of our numerical result. However, it is important to understand that such a limit is defined in the context of a disturbed two-body problem (i.e. considering Galactic perturbations) and we cannot directly compare this limit with the isolated two-body problem. Then, the key to our analysis is that the characteristic scale define by $r_J$ is obtained from the Jacobi constant, which is an integral of motion of the problem defined by
\begin{equation}
E_J = \dfrac{\rm{v}^2}{2} - \frac{\mu}{r} - 4 A_g \Omega_g x^2 + \nu_g^2 z^2   \,  \rm{,} \label{eq5}
\end{equation}
where v$^2 = \dot{x}^2 + \dot{y}^2 + \dot{z}^2$. The two first terms only depend on the semi-major axis and correspond to the energy of an isolated two-body problem (i.e. $- \, 0.5 \, \mu \, a^{-1}$), and the two last terms depend on all orbital elements and correspond to the tidal influence of the Galactic potential. Then, for larger separations the two tidal terms in eq. (\ref{eq5}) become important, and unlike an isolated two-body problem (i.e. set \textbf{c1b}) the rate of disruption of the WBS in the set \textbf{c1} depends not only on the semi-major axis, but also on the other orbital elements.

The Jacobi constant can be separated into two parts: first, the terms corresponding to the two-body problem, which represent the energy of the WBS in the sample \textbf{c1b}; and  second, the terms of the Galactic potential, which have influence on the disruption process of the set \textbf{c1}. In particular the third term in eq. (\ref{eq5}) is negative similar to the energy of the two-body problem, and it can be considered as a protection produced by the tide forces  against the disruption of the pair by an external perturbation, such as a stellar passage. So, it is possible to predict that more energy is necessary to break up the pair because of this term. On the other side, the last term has the opposite effect because it is positive, which reduces the bond energy of the pair and increases the probability of disruption. It is worth noting the dependence of the last term on the inclination of the WBS orbit.

Our numerical study (Fig. \ref{fig1}, bottom panel) shows, for the \textbf{c1b} set, that number of disrupted systems does not depend on the inclination, while for the \textbf{c1} set the probability of disruption reaches a maximum for $I \, = \, \pm $ 90$^\circ$. Such results are in agreement with our analytical predictions and demonstrate that the inclusion of the Galactic potential reduces the random effect of stellar passages. For WBS with the orbital plane close to the Galactic plane the $z$-component is close to $0$, so the tide effect (third term of eq. (\ref{eq5})) increases the bond energy of the pair and the stellar encounters need to transfer more energy to break up the binary star. Instead, for highly inclined orbits we have the opposite case; now the $x$-component is close to $0$ and the Galactic potential (fourth term of eq. (\ref{eq5})) reduces the  energy required to disrupt the stellar pair.

 \begin{figure}
\centering
\subfigure{\includegraphics[width=0.7\textwidth]{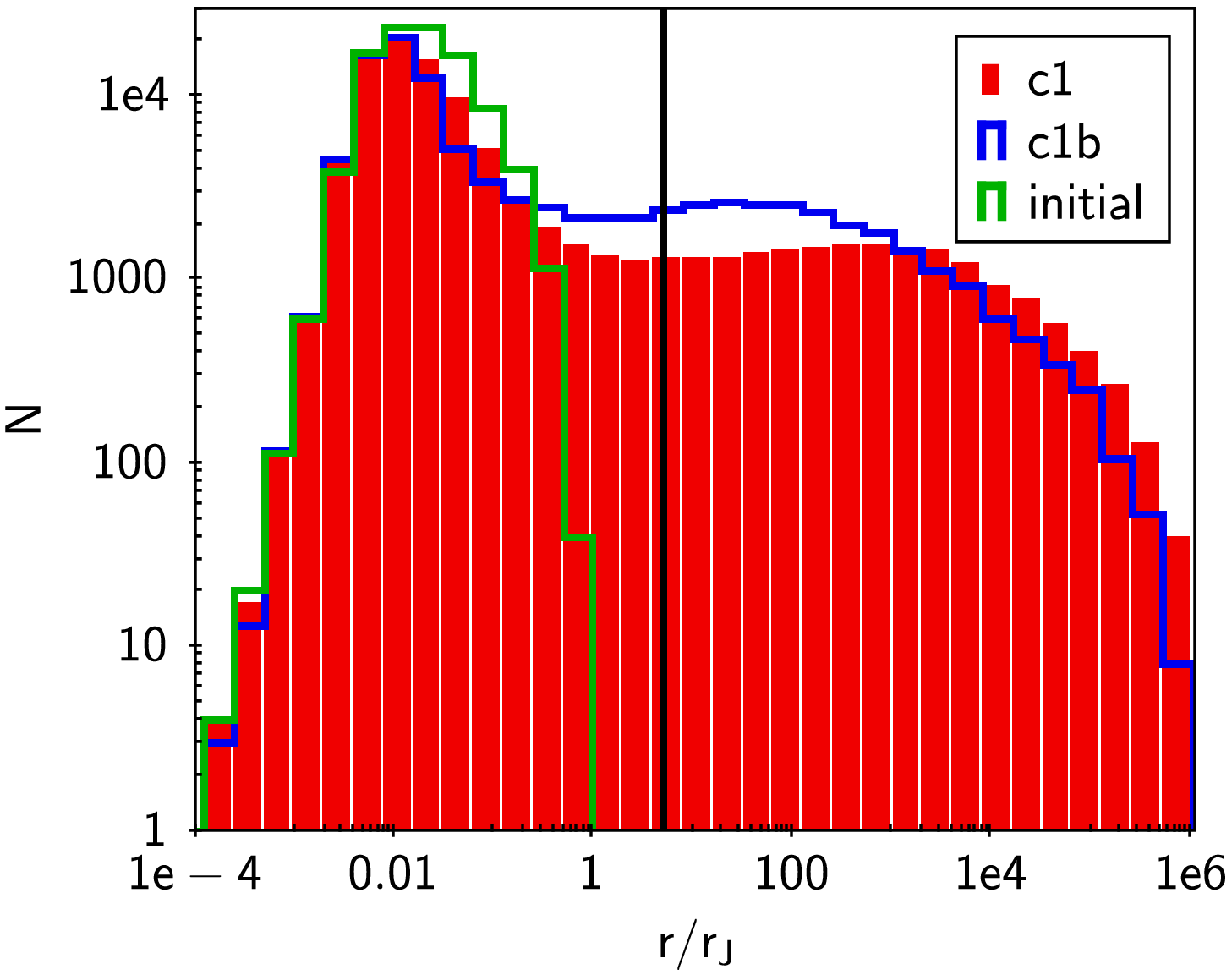}}
\subfigure{\includegraphics[width=0.7\textwidth]{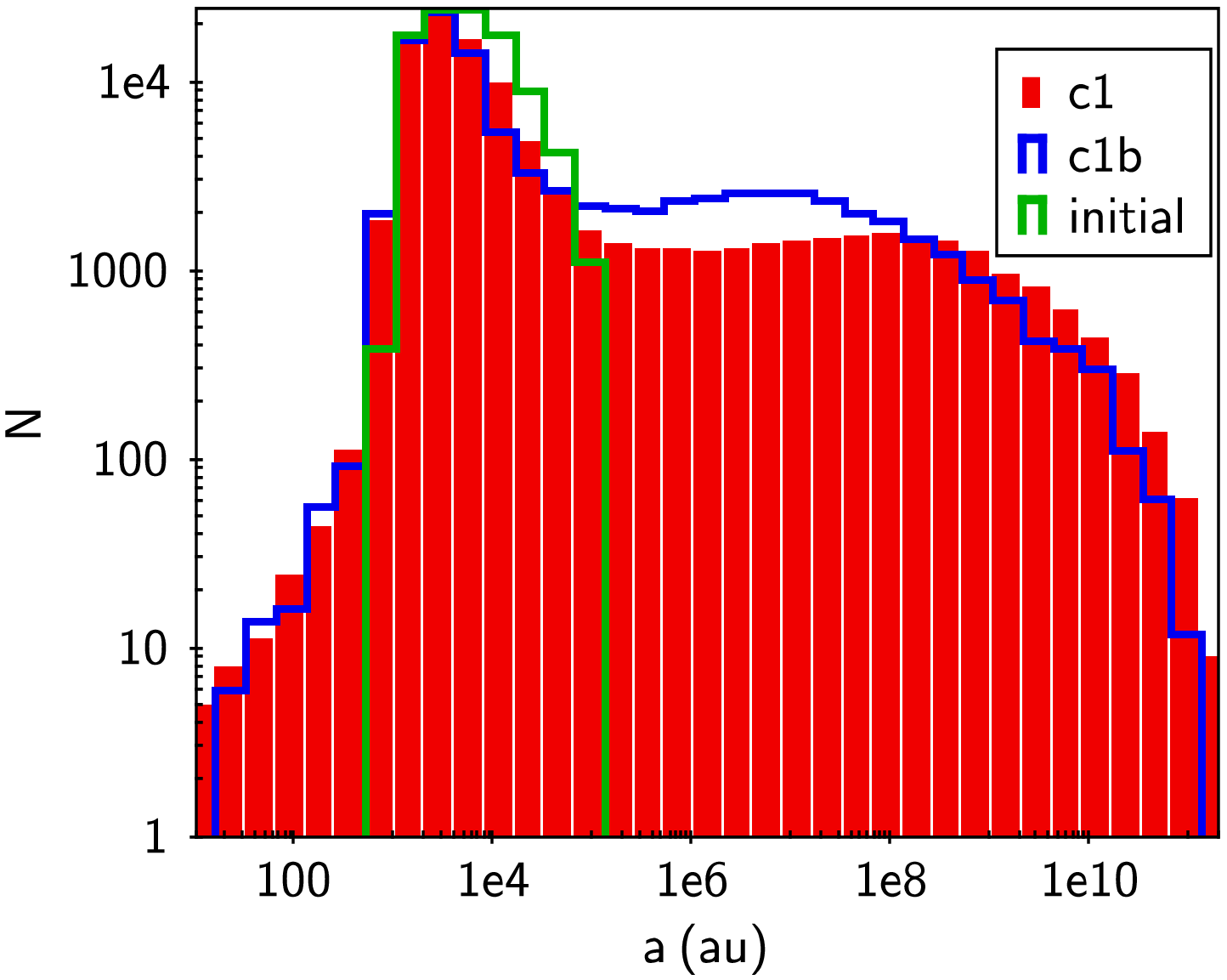}}
\caption{{\small Final distribution for the sets \textbf{c1} (red), and \textbf{c1b} (open blue) after $T=$ 10 Gyr of evolution. In green we include the histogram of the initial distribution of both samples. The top and bottom panels correspond to the final separation divided by $r_J$ and the final semi-major axis, respectively. The black vertical line in the top graph corresponds to $\sim$ 5 $r_J$.}}
\label{fig2b}
\end{figure}

\cite{jiang10} found numerically for a sample similar to our \textbf{c1} set that the influence of all the orbital elements in the disruption process can be represent by the separation of the stellar pair $r$, scaled by $r_J$. In Figure \ref{fig2b} we plot the distributions of the final separation ($r$) and the final semi-major axis ($a$) for all the WBS of each sample, even the disrupted systems, where the histogram for the \textbf{c1} set is shown in red and that for the \textbf{c1b} set in open blue, while the initial distribution is indicated with a green histogram. In the top panel of  Fig. \ref{fig2b} the distribution of $r/r_J$ is shown, and for the \textbf{c1} set our results agree with the prediction of Jiang \& Tremaine (2010). There is a second maximum or exterior peak beyond 100 $r_J$ that corresponds to the disrupted WBS, and there is a minimum at $\sim$ 5 $r_J$, which is indicated by a black line in the graph. On the other hand, for the \textbf{c1b} set the second peak is higher than the maximum observed for the \textbf{c1} set but it is below 100 $r_J$, while the minimum is close to $\sim$ 1 $r_J$. 
On the bottom panel of Fig. \ref{fig2b} we can see that the distribution of the semi-major axis has the same structure of two peaks as the top panel. If we approximate the evolution of a WBS in the Galaxy by a perturbed two-body problem, in a first approximation the semi-major axis can be considered as indicative of the energy of the system. Therefore, the shift of the exterior peak for the \textbf{c1} set seems to confirm that a WBS disturbed by the Galaxy must take more energy from the passing stars to be disrupted.

\begin{figure}
\centering
\subfigure{\includegraphics[width=0.7\textwidth]{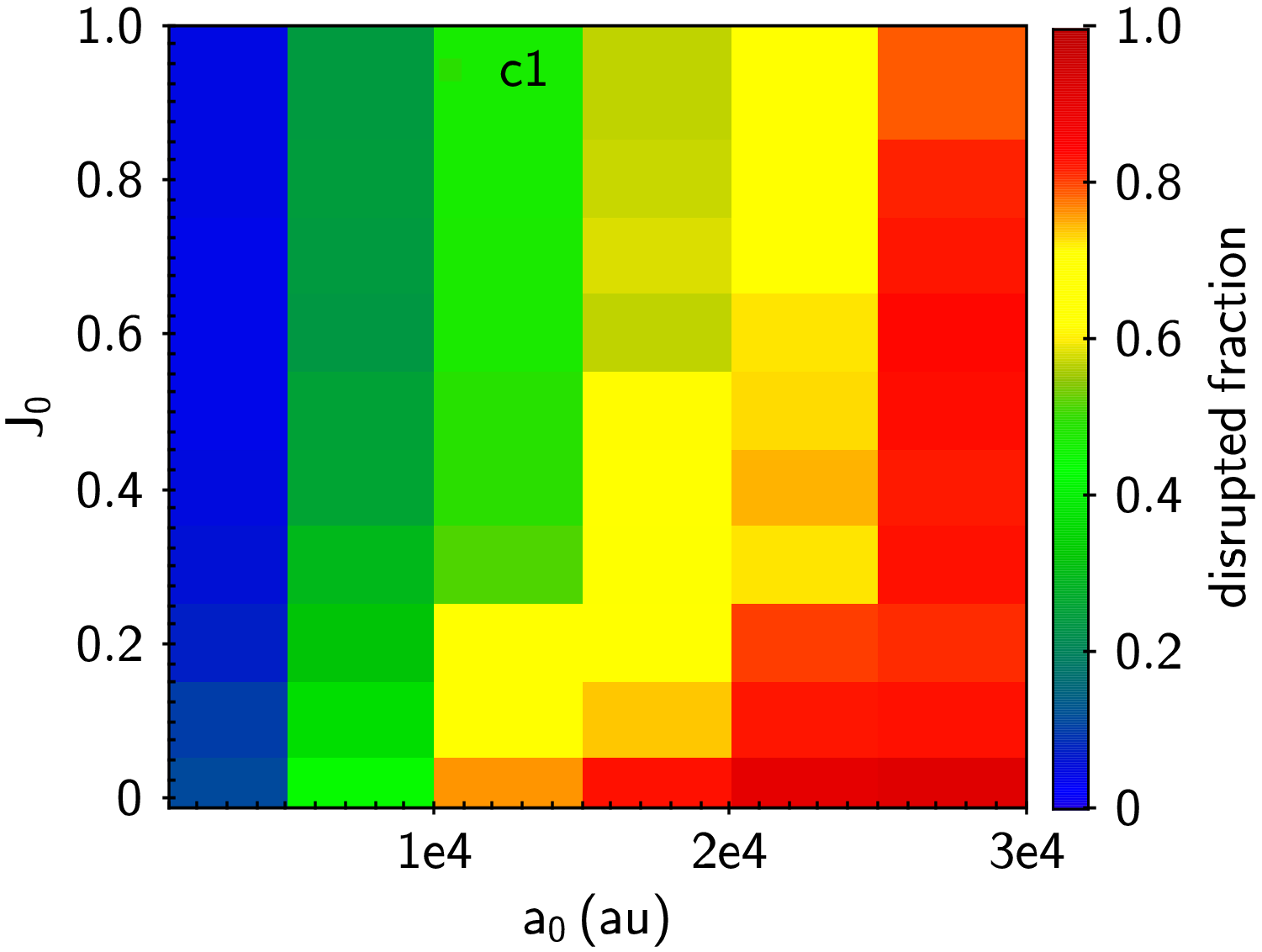}}
\subfigure{\includegraphics[width=0.7\textwidth]{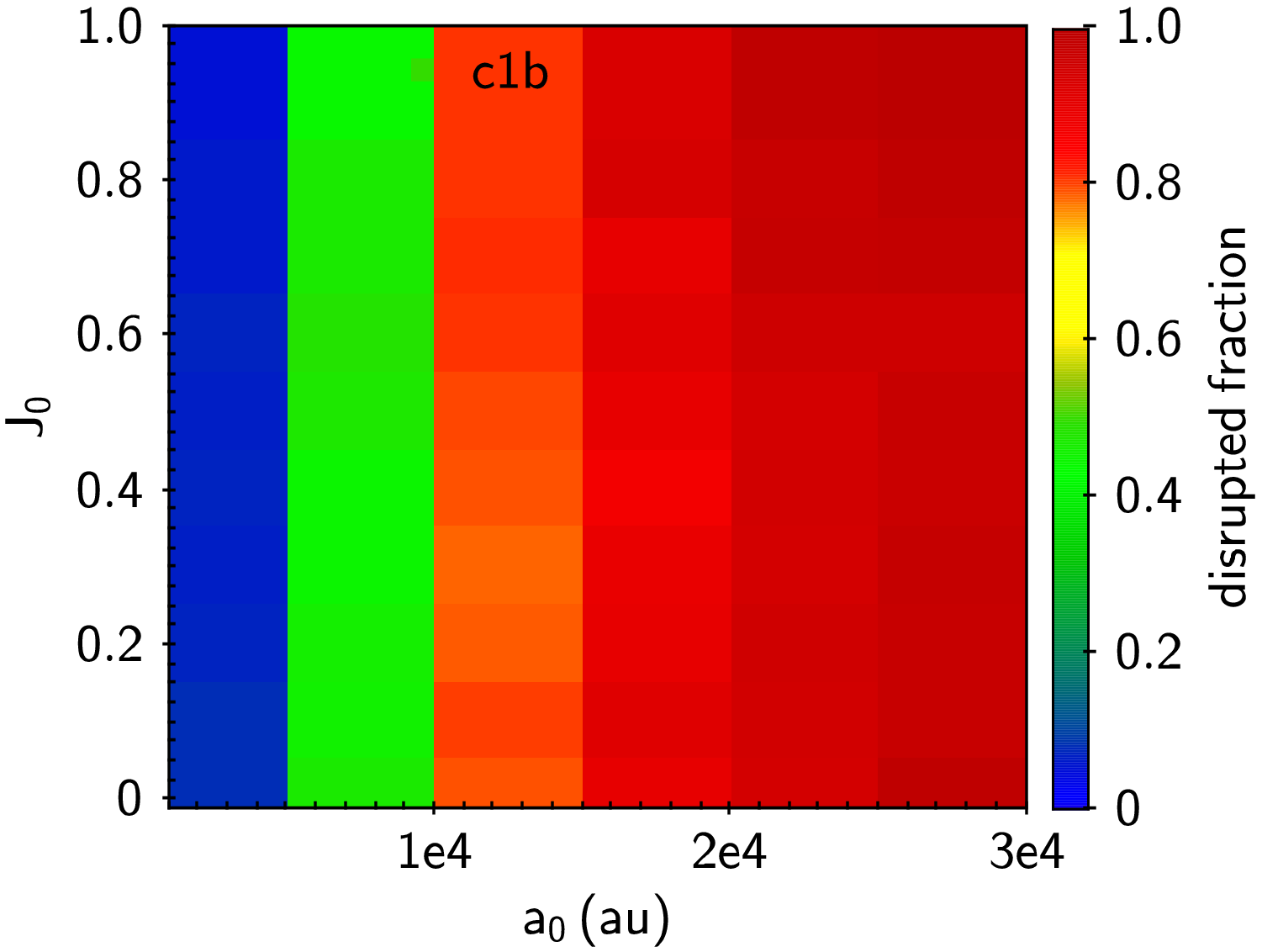}}
\caption{{\small  Fraction of disrupted WBS as a function of $a_0$ and  $J_0$. The top and bottom panels correspond to the sets \textbf{c1} and \textbf{c1b}, respectively. We can see a dependence of the  \textbf{c1} set on the initial dimensionless projection of the angular moment because for small values of $J_0$ the disruption rate increases.}}
\label{figqb}
\end{figure}

In order to show the dependence of the bond energy with the orbital elements for the \textbf{c1} set we considered the works of Heisler \& Tremaine (1986) and Correa-otto et al. (2017), who found that the phase space of a binary star in the Galaxy is parameterized by the semi-major axis and the dimensionless projection of the angular moment ($J= \sqrt{1-e^2} \cos{I}$). From Figures 9 and 10 of Correa-otto et al. (2017) we can deduce that for a WBS with high values of $J$ the extension of the phase space decrease and it can not reach high values of $e$. So, just as we deduced in our previous analysis, it is possible to predict that in the case of a sample affected by tidal effects (\textbf{c1}) the stellar passages have to transfer more energy to break up the pair in comparison with the other sample (\textbf{c1b}) because the encounters have to change not only the semi-major axis but also the parameter $J$. Figure \ref{figqb} shows the disrupted fraction as function of the initial semi-major axis ($a_0$) and the initial dimensionless projection of the angular moment ($J_0$) in a grid of 6 by 10 bins, and we can see that for the \textbf{c1} set (top panel) the process of rupture depends on the parameter $J_0$ (i.e. $e_0$ and $I_0$). This result agrees with our theoretical prediction that when the value of $J_0$ increases, the disruption probability decreases. For small values of $J_0$ where the phase space extends to $e \sim$ 1 (i.e. $J_0 < 0.1$), the dependence of the disrupted fraction on the semi-major axis is similar for both samples.

Finally, we can describe the effect of the Galactic potential as a protection mechanism for the WBS against the random disruption effects of the stellar passages. In the dynamic evolution of a sample of WBS in the solar neighbourhood, each binary star can gain or lose energy through stellar passages; successive encounters can reduce the bond energy until the stellar pair breaks up, but the tidal secular influence of the Galactic potential increases such bond energy in some direction. Then, the final result is a lower probability of dissociation for the sample with respect to the case without tidal influence.

\subsection{Dynamical consequences on the planetary region}

 \begin{figure}
\centering
\subfigure{\includegraphics[width=0.7\textwidth]{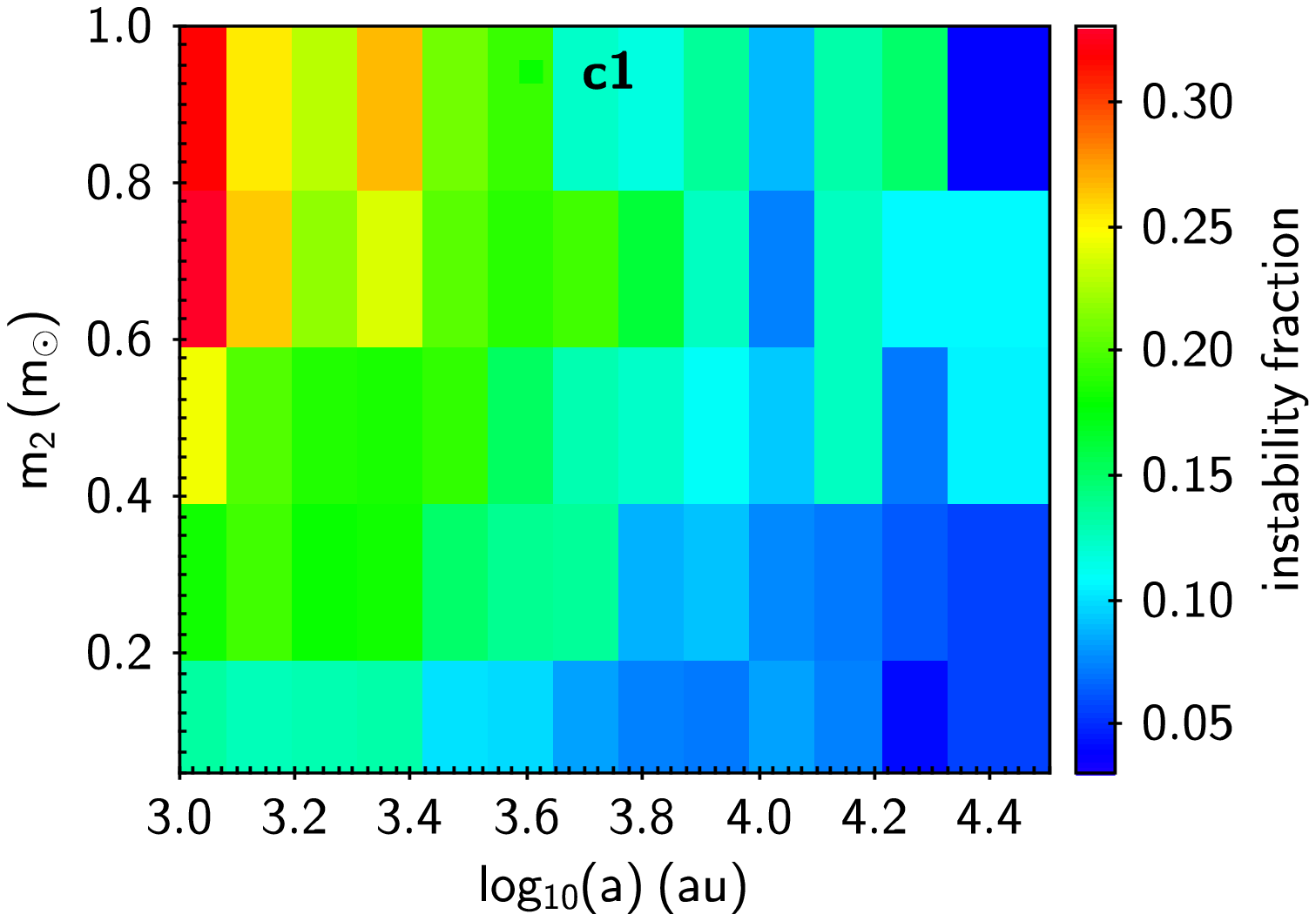}}
\subfigure{\includegraphics[width=0.7\textwidth]{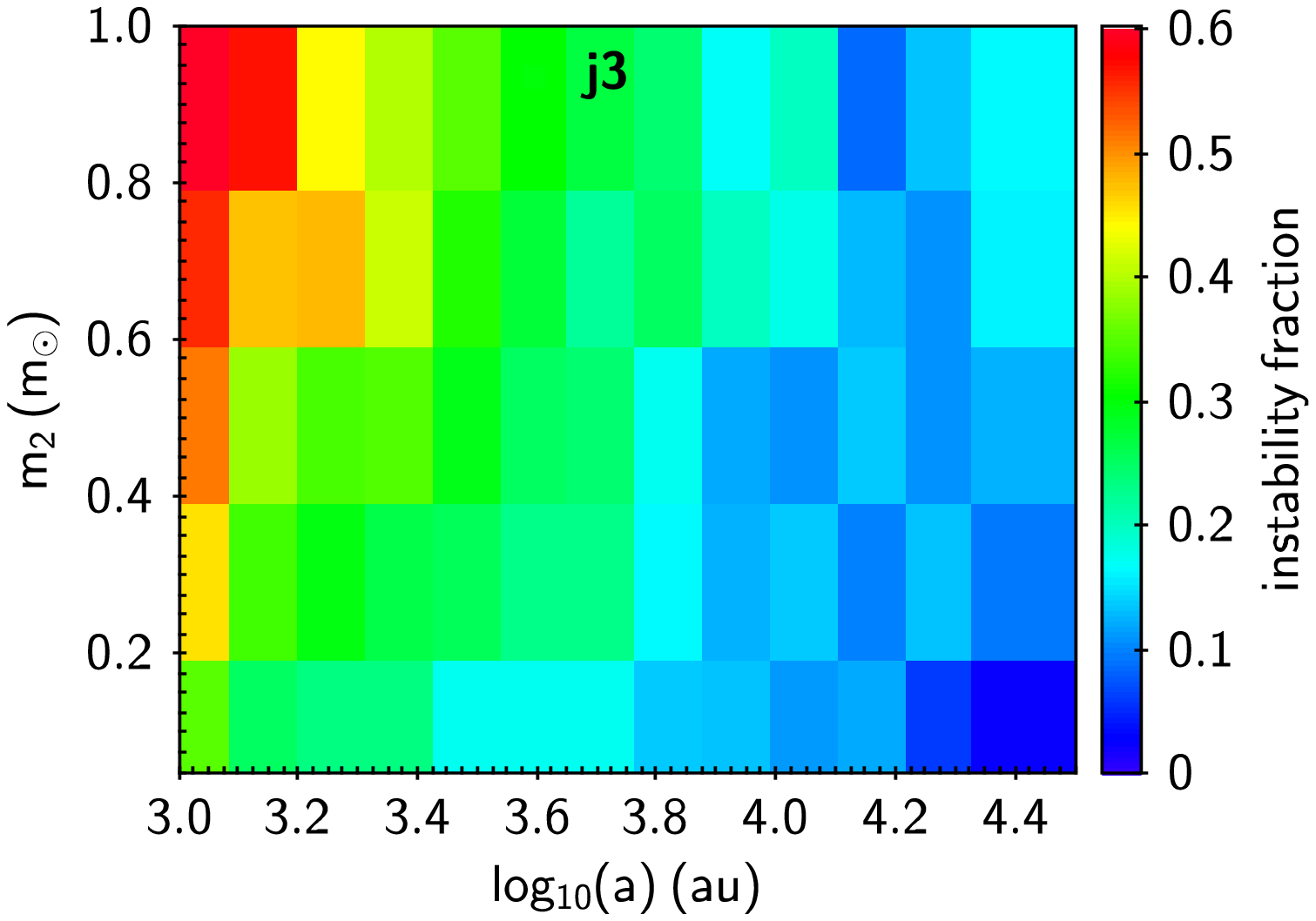}}
\caption{{\small Percentage of WBS in the sets \textbf{c1} and \textbf{j3}  with the planetary region around $m_1$ disturbed by the secondary star, as a function of the semi-major axis ($a$) and secondary mass ($m_2$). The maps show a good agreement with the results of \cite{kaib13}. }}
\label{fig5}
\end{figure}

In our simulations we do not include planets, but it is possible to estimate the disturbing effects of the Galaxy in the planetary region around $m_1$. The main external phenomena considered in this work are the tidal field of the Milky Way and passing field stars, which act indirectly in the area of interest. The first case corresponds to a long duration effect and takes place in the interval between stellar passages. During this stage the planetary region can be disturbed by $m_2$ if the secondary stars occasionally reaches a small distance of pericentre (Kaib \& Raymond 2014; Correa-Otto et al. 2017). In the second case, the disturbance by an encounter is a fast phenomena in which the gravitational influence of $m_2$ and/or $m_3$ is defined by a fortuitous close approach of each one of these objects to the main star.

Most of the known exoplanets have their orbits with $a<$ 5 au, and the same is true for the exoplanets in WBS (last column in Table \ref{table0}). However, in our solar system the planetary region extends until 30 au, and \cite{kaib13} mentioned that we are probably observing the most inner planets of the exoplanetary systems due to the methods of detection. Then, if we suppose a planetary system similar to the solar system orbiting the star $m_1$ of the binary, the extent of the gravitational influence of the secondary on the  planetary region can be estimate by the critical periastron ($q_C$)  defined in Kaib et al. (2013).

Thus, for a wide binary star we consider that the hypothetical planetary system is destabilized when in the numerical integration the minimal distance between $m_1$ and $m_2$ evolves below the critical distance $q_C$, at least one time. The results of  Kaib et al. (2013) were obtained for a main star of one solar mass and more than 70 \% of our sets of WBS have  primary stars with masses $\sim$ 1 M$_\odot$; this makes it possible to compare our results with those of Kaib et al. (2013).

Table \ref{table1} shows the percentage of planetary regions perturbed by $m_2$ for each set and subset. We can see a difference between the percentage obtained for the \textbf{j3} set and the other two sets. Since the main difference between these sets is a different initial distribution of $e$ for \textbf{j3}, we can deduce that this orbital element is responsible for the planetary dynamics in WBS. On the other hand, the disturbed fraction in the \textbf{c1a} subset is twice that of \textbf{c1b} and is close to that of \textbf{c1}. This seems to indicate that the Galactic potential is the main effect that modulates the process of planetary destabilization.

We also reproduce the dynamical map presented in Fig. 2 (a) of Kaib et al. (2013) using our samples \textbf{c1} and \textbf{j3}. We divided each initial sample in a grid of 13 by 5 bins in semi-major axis (in log-scale) and mass of the companion, and we calculated the total number of WBS in each bin with their hypothetical planetary system destabilized and divided this quantity into the total number of systems in that bin. Figure \ref{fig5} shows our results, where we can see values in a range of 5-30 \% for the set with uniform distribution of $e$ and values in a range of 5-55 \% for the set with the initial distribution of eccentricity  similar to that of Kaib et al. (2013) (i.e. $e^2$). The percentages shown for the set \textbf{j3} are similar to those of Kaib et al. (2013), which confirm   the efficiency of the method of critical periastron ($q_C$) and  the outer influence of the Galaxy as an important source of perturbation for the planetary systems in WBS.  

The dynamical structure of our maps is similar to that of the map shown by Kaib et al. (2013) despite the difference in the initial distributions of semi-major axis and secondary mass. Therefore, from our results we can say that the initial distribution of the eccentricity and its subsequent evolution, which is affected mainly by the secular effect of the Galactic potential Heisler \& Tremaine (1986) and Correa-otto et al. (2017), modulates the planetary stability in WBS. 

\begin{table}
\caption{Percentage of WBS with the planetary region around $m_1$ disturbed by $m_2$ for all the sets. }
\label{table1}
\centering
\begin{tabular}{c c }
\hline\hline
sample &  disturbed    \\
       &  fraction    \\ 
 \hline\hline
c1 & 0.16  \\
c1a & 0.11  \\
c1b & 0.06  \\
k2 & 0.15  \\
j3 & 0.27  \\ 
\hline
\end{tabular}
\end{table}

Finally, we analyse the direct effect of the third star on the planetary region. Following Laughlin \& Adams (2000), we estimate the change of eccentricity in a hypothetical Neptune, at 30 au with $e_N=$ 0.01. Considering that the close approach of the third star to $m_1$ produces an instant force, we estimate the $\Delta e $ in the orbit of our hypothetical Neptune in the context of a perturbed two-body problem (Murray \& Dermott 1999). We found a probability smaller than 0.5 \% for a change $\Delta e  > 0.01$ on the three sets. Then, for statistical studies we can ignore the direct influence of the third star in the planetary region.

\section{Conclusions}\label{conclu}

In this paper, we presented a statistical study about the temporal evolution of a synthetic set of binary star systems in the solar neighbourhood with orbital and physical characteristic similar to the population of wide binary stars with exoplanets (Table \ref{table0}) and taking into account the effects produced by the tidal field of the Milky Way and the perturbations produced by passing stars. We carry out simulations using three sets of $10^5$ WBS with different initial distributions of $a$ and $e$.  The dynamic evolution of the WBS in each set was followed by solving the exact numerical equations of motion during 10 Gyr, i.e. the approximated age of the thin disk.

Our results show that the three sets of WBS have similar final orbital element distributions regardless their initial configuration. Therefore, we conclude that the external effects modified the samples and makes them converge to a \textit{standard configuration} during the temporal evolution. The most important phenomenon that alters a population of WBS is the disruption of binary pairs, and this effect is independent of the masses of the pair. Thus, for the population of WBS studied the main characteristics of its final \textit{standard configuration} are
\begin{itemize}
\item There is an accumulation of systems within $a \in (2000,5000)$ au.
\item The population is dynamically hot  (i.e. high eccentricities).
\item There is a gap in the distribution of WBS  in a direction perpendicular to the Galactic plane (i.e. $I \sim \pm 90^\circ$ ).
\item The final mass distribution preserves the original form.
 \end{itemize}
A denser galactic environment could be more efficient at generating a \textit{standard configuration}, for example by a faster disruption process on the most separated WBS.

From our numerical results and the analytical interpretation of these results, we conclude that the disruption process is dominated by the cumulative effect of stellar passages with the Galactic potential acting like a protection mechanism against the break up of the stellar pair.  We found that the most important characteristics of the WBS that regulates this phenomenon are first, the separation of the pair ($r$), which has been demonstrated by Jiang
\& Tremaine (2010); and second, its orientation with respect to the Galactic plane ($I$). This last point predicts a lack of binary systems with orbits perpendicular to the Galactic plane in the solar neighbourhood, which agree with the predictions of our analytical study: a stellar pair in an orbit with high inclination has a smaller bond energy than those in a coplanar orbit.

Thus, as the population of WBS with exoplanets reaches a \textit{standard configuration} as a result of the Galactic effects, we can get rid of the dynamical problem of the definition of the initial distributions for the orbital elements. Our results strengthen the statistical study of Kaib et al. (2013) because the 2600 combinations of WBS considered in that work (100 times less than our sample) are enough to reach correct results.  However, our results imply a problem for the formation studies of WBS (e.g. Tokovinin (2017)) because the current distribution of the orbital elements would not show traces of the initial distribution.

Additionally, we found an agreement between our results and those of Kaib et al. (2013) for the percentage of disrupted WBS. \textbf{The importance of this agreement is highlighted when we consider that we solved the complete equations of motion, while they applied the model of impulse approximation.} Although this model has been tested in the case of the restricted three-body problem (Yabushita et al. 1982; Scholl et al. 1982; Dybczynski 1994; Eggers \& Woolfson 1996), there are no studies about its accuracy for a general three-body problem. Thus, we conclude that for a dynamical analysis about the evolution of the population of WBS with exoplanets in the Galaxy, the impulsive hypothesis gives reliable statistical results.

Finally, we applied the criteria of critical periastron proposed by Kaib et al. (2013) to estimate the indirect influence of the Galactic environment on the planetary region around the main star of a WBS. Our results show that the effects of the external perturbations considered in this work are in agreement with that obtained by Kaib et al. (2013). However, we find that different initial configurations of the samples of WBS produce varying levels of dynamic excitation in the planetary system, with the initial distribution of eccentricity of the secondary component as the main modulator of the problem. Therefore, from our partial analysis (i.e. without planetary systems), we concluded that more studies about the topic of exoplanets in WBS are needed to improve our understanding of the dynamical influence of the secondary star orbit on the planetary region around the main star.

\section{Bibliography}

\begin{itemize}
\item Andrade-Ines, E. \& Michtchenko, T. A. 2014, MNRAS, 444, 2167
\item Bahcall, J. N., Hut, P., \& Tremaine, S. 1985, AJ, 290, 15
\item  Binney, J. \& Tremaine, S. 2008, Galactic Dynamics: Second Edition (Princeton University Press)
\item Brunini, A. 1995, A\&A, 293, 935
\item Brunini, A. \& Fernández, J. 1996, A\&A, 308, 988
\item Correa-Otto, J., Calandra, F., \& Gil-Hutton, R. 2017, A\&A, 600, 59
\item Duncan, M., Quinn, T., \& Tremaine, S. 1987, AJ, 94, 1330
\item Dybczynski, P. A. 1994, CeMDA, 58, 139
\item Eggers, S. \& Woolfson, M. 1996, MNRAS, 282, 13
\item Fouchard, M., Froeschlé, C., Valsecchi, G., \& Rickman, H. 2006, CeMDA, 95,
299
\item García-Sánchez, J., Weissman, P., Preston, R., \& et al. 2001, A\&A, 379, 634
\item Heggie, D. C. 1975, MNRAS, 173, 729
\item Heggie, D. C. 2001, in The Restless Universe, ed. B. A. Steves \& A. J. Maciejew-
ski, 109–128
\item Heisler, J. \& Tremaine, S. 1986, ICARUS, 65, 13
\item Holman, M. J. \& Wiegert, P. A. 1999, AJ, 117, 621
\item Jiang, Y. F. \& Tremaine, S. 2010, MNRAS, 401, 977
\item Kaib, N. A. \& Raymond, S. N. 2014, ApJ, 782, 60
\item Kaib, N. A., Raymond, S. N., \& Duncan, M. 2013, Nature, 493, 381
\item Kaib, N. A., Roskar, R., \& Quinn, T. 2011, Icarus, 215, 491
\item Kennedy, G. M., Matrá, L., Marmier, M., et al. 2015, MNRAS, 449, 3121
\item Laughlin, G. \& Adams, F. C. 2000, Icarus, 145, 614
\item Levison, H. F. \& Dones, L. 2001, AJ, 121, 2253
\item Moro-Martín, A., Marshall, J. P., Kennedy, G., et al. 2015, ApJ, 801, 143
\item Murray, C. D. \& Dermott, S. F. 1999, Solar system dynamics (Cambridge Uni-
versity Press)
\item Ninkovic, S. \& Trajkovska, V. 2006, Serb. Astron. J., 172, 17
\item Rabl, G. \& Dvorak, R. 1988, A\&A, 191, 385
\item Reid, I. N., Gizis, J. E., \& Hawley, S. L. 2002, AJ, 124, 2721
\item Rickman, H. 1976, BAICz, 27, 92R
\item Rickman, H., Fouchard, M., Froeschlé, C., \& Valsecchi, G. B. 2008, CeMDA, 102, 111
\item Roell, T., Neuhauser, R., Seifahrt, A., \& Mugrauer, M. 2012, A\&A, 542, A92
\item Roskar, R., Debattista, V. P., Quinn, T. R., Stinson, G. S., \& Wadsley, J. 2008, ApJ, 684, 79
\item Scholl, H., Cazenave, A., \& Brahici, A. 1982, A\&A, 112, 157
\item Sellwood, J. A. \& Binney, J. J. 2002, MNRAS, 336, 785
\item Tokovinin, A. 2017, MNRAS, 000, 000
\item Yabushita, S., Hasegawa, I., \& Kobayashi, K. 1982, MNRAS, 200, 661
\item Zakamska, N. L. \& Tremaine, S. 2004, AJ, 128, 869
\end{itemize}

\section*{Acknowledgements}

The authors gratefully acknowledge financial support by CONICET through PIP 112-201501-00525. The authors are grateful to the anonymous referee for numerous suggestions and corrections on this paper.





\end{document}